\documentclass[prb,a4paper,showpacs,twocolumn,superscriptaddress,longbibliography]{revtex4-1}

\usepackage{textcomp}
\usepackage{amssymb}
\usepackage{amsmath}
\usepackage{amsfonts}
\usepackage{graphicx}
\usepackage{bm}
\usepackage{xcolor}
\usepackage{multirow}
\usepackage{natbib}
\usepackage{hyperref}
\usepackage{mathrsfs}

\begin{document}

\title{Topological numbers of quantum superpositions of topologically non-trivial bands}

\author{E. V. Repin}

\affiliation{Kavli Institute of Nanoscience, Delft University of Technology, 2628 CJ Delft, The Netherlands}

\author{Y. V. Nazarov}

\affiliation{Kavli Institute of Nanoscience, Delft University of Technology, 2628 CJ Delft, The Netherlands}

\begin{abstract}
Topological properties of the wavefunction manifolds - bands are in focus of modern condensed matter research. In this Article, we address the definition and values of topological numbers of quantum superpositions of the topologically distinct bands. The problem, although simple in essence, can be formulated as a paradox: it may seem that quantum superposition implies non-integer topological numbers.

We show that the results are different for superpositions that are created dynamically and for those obtained by stationary mixing of the bands. For dynamical superpositions, we have found that an observable commonly witnessing a topological number is non-integer indeed. For static superpositions, the resulting bands retain integer topological numbers. We illustrate how the quantization of topological number is restored upon avoided crossing of topologically distinct subbands. The band crossings may result in the exchange of topological numbers between the bands upon changing the parameters describing the bandstructure. This is a phase transition between the phases defined as sequences of topological numbers of the bands. We consider complex phase diagrams arising in this context and show the absence of triple critical points and abundance of quadruple critical points that are rare in common phase diagrams. We illustrate these features with a bilayer Haldane model.
\end{abstract}

\maketitle

\section{Introduction}
The notion of the topology of the wavefunction manifolds has been being discussed for many years (see e.g. Refs.\cite{PhysRevB.31.3372,Berry45}). In addition to the topological applications in the fields such as cosmology\cite{cosm}, quantum field theory\cite{witten}, classical integrable Hamiltonian dynamics\cite{fomenko}, etc. it has also been understood that the topology may play an important role in condensed matter\cite{ryu}, in particular in the quantum description of crystal solids\cite{Qi,kane}. The Chern insulator\cite{PhysRevB.74.235111} is an example of such an application. Let us describe the simplest case of a 2-dimensional Chern insulator. The Hamiltonian of this periodic solid is a matrix defined in the compact space of two quasimomenta $q_{\alpha}$, $\alpha=1,2$ and is continuous. The eigenbasis of the Hamiltonian is also parameter-dependent and the resulting manifolds of parameter-dependent wavefunctions that belong to a certain eigenvalue are usually called bands. It is important to note that continuity of the Hamiltonian does not immediately imply the continuity of these wavefunctions. Since the eigenfunctions are defined upon the phase factor $e^{i\chi(\vec{q})}$, they have to be continuous only upon a phase factor. Let us call this property a quasi-continuity.

Mathematically a band can be regarded as a section of a 1-dimensional linear bundle that can be characterized by an integer Chern number in a standard way\cite{Berry45,Nakahara}. It is common to define the Berry curvature of the band $k$ is commonly defined as\cite{Berry45} $B_{\alpha \beta}^{(k)}=-2{\rm Im}\langle \partial_{q_\alpha} \psi_k(\vec{q})|\partial_{q_\beta}\psi_k(\vec{q})\rangle$. The first Chern number is an integral of the Berry curvature over the compact space of $\vec{q}$ and has to reduce to an integer times $(2\pi)$. One can also define the Hamiltonian in the space of $\vec{q}$ with more dimensions. Then the first Chern number is defined as the integral over any 2-dimensional compact subspace of $\vec{q}$. 

In physical terms first Chern number can be directly related to the transconductance of the system. To establish this, one may utilize the description in terms of the semi-classical equations of motion. As known\cite{PhysRevB.59.14915}, a nonzero value of the Berry curvature brings about a nonzero drift velocity transverse to the external force $F^\beta$. Average of this drift velocity over the $\vec{q}$-space is expressed in terms of the first Chern number. This explains the transverse conductance quantization in the QHE setups\cite{qhe,PhysRevB.31.3372}. We note that the above consideration implies that the initial wavefunction is the eigenstate of the Hamiltonian. It is not evident if the conclusion is valid for more complex initial wavefunctions.

The linearity is one of the basic postulates of quantum mechanics: a linear superposition of two wavefunctions is also a valid wavefunction of the system. 
What are the topological properties of the superposition of topologically distinct bands? To comprehend why the question is not trivial let us consider a short example. Let us take two bands of quasi-continuous wavefunctions $|\psi_{0}(\vec{q})\rangle$, $|\psi_{1}(\vec{q})\rangle$ with different first Chern numbers $C_0=0$ and $C_1=1$. 
Let us consider a superposition of those bands with parameter-independent coefficients $a,b\ne 0$
\begin{equation}
|\psi(\vec{q})\rangle=a|\psi_0(\vec{q})\rangle+b|\psi_1(\vec{q})\rangle
\label{super}
\end{equation}
What is the Chern number of this superposition? It is a weighted sum of two Chern numbers, $|b|^2$ which is generally not integer. The topology dictates that the Chern number must be integer\cite{Nakahara}. So, this presents an apparent paradox. 

The solution is simple: the superposition of two quasi-continuous wavefunctions is not quasi-continuous and is not subject to topological classification. The superposition can be made quasi-continuous by choosing the proper parameter dependence of $a$ and $b$. This restores the quantization of Chern number.

This sets the goal of this Article: to investigate the topological properties of superpositions. We address the superpositions of two kinds. A ${\it dynamic}$ superposition is created at a given value of $\vec{q}$ and then evolves in accordance with the Hamiltonian dynamics. A ${\it static}$ superposition is obtained by modification of the stationary Hamiltonian that mixes the bands. Let us shortly describe the results. Firstly, we consider dynamic superpositions. We investigate the time evolution of a particle in the Chern insulator that is prepared initially in a superposition state. In such insulators the transverse current response - the transconductance is supposed to witness the Chern number. The quantization of transconductance is related to the quantization of transmobility that gives a transverse velocity. We find for a superposition that the resulting transmobility is indeed proportional to the weighted sum of Chern numbers, so it is not a subject of topological quantization. This can be deduced that the wavefunction is not periodic in time and its time-dependence is complex. Although dynamics are equivalent to the slow change of parameters $\vec{q}$ in the Hamiltonian the wavefunction does not remain quasi-continuous on a closed trajectory in a parameter space. This is different from the dynamics of the eigenfunctions. Thus the resulting wavefunction is not a subject of topological classification.
 
Next, we investigate the static superpositions, made by adding the mixing matrix elements between the bands into the Hamiltonian. We investigate the topological properties of the resulting eigen-bands. These eigen-bands are quasi-continuous and we establish the topological restriction on the parametric dependence of the mixing matrix element: it must be zero at least in one point in $\vec{q}$-space. The Chern number of a static superposition is thus integer and may change abruptly upon changing of the parameters of the bandstructure. Such an abrupt change is a topological phase transition, so this naturally brings us to the consideration of possible phase diagrams. The phases we consider are defined by a set of first Chern numbers attributed to each band with Chern numbers being ordered with increasing energy of the bands. We investigate the critical points in these phase diagrams and find no triple points. The critical points are quadruple connecting 4 regions in 2d parameter space. There are two kinds of quadruple points with either 4 or 3 different phases in the adjacent regions. This is different from the case of phase diagrams for usual phase transitions where generically triple points are present\cite{vol5}. We extensively illustrate these features of phase diagrams with a specific example of a bilayer Haldane model.

The paper is organized as follows. In Sec. \ref{Sec:evol} we derive the value of the transconductance of a particle prepared initially in the superposition state. The discussion of the restrictions imposed by general topological considerations on the mixing matrix element between two topological bands is given in Sec. \ref{Sec:propt}. In Sec. \ref{Sec:static} we address the properties of the topological numbers exchange for between two static superpositions. In Sec. \ref{Sec:moar} we investigate the case of multiple bands introducing and discussing the general features of the phase diagrams. In Sec. \ref{Sec:haldane} we illustrate these general features inspecting the phase diagrams of the topological phase transitions in the bilayer Haldane model. We conclude in Sec. \ref{Sec:Sum}. The present a note on a specific degenerate case of the bilayer Haldane model in App.$\ref{Sec:App1}$.

\section{Adiabatic evolution of the superposition\label{Sec:evol}}
In this Section we consider the adiabatic evolution of a particle initially prepared in a superposition of two states that belong to different bands. We will compute the transmobility $\mu$ of a particle defined as the proportionality coefficient between the drift velocity in the direction perpendicular to the external force and the external force, $v_\alpha=\mu e_{\alpha \beta}F_\beta$. As we will show below, for a particle in a certain band
\begin{equation}
\mu=-\frac{2\pi C}{\hbar \Omega}
\label{mu}
\end{equation}
where $C$ is an integer Chern number of the band and is the volume of the Brillouin zone $\Omega=\int dq_1 dq_2$. The transverse current density of a many-body system at zero temperature is a sum over filled bands labeled by $j$
\begin{equation}
j_\alpha=e\sum_{j}^{}v_\alpha^j n^j=e^2 \sum_{j}^{}\mu_j n^j \epsilon_{\alpha \beta} E^\beta
\end{equation}
where $\mu_j$ is the transmobility in the band $j$ and the particle density $n_j=\frac{\Omega}{(2\pi)^2}$ in a filled band does not depend on a band . With this, the transconductance\cite{qnbook}, the proportionality coefficient between the transverse current and the voltage $I_x=G_{xy} V_y$ is quantized in the units of $e^2/(2\pi \hbar)$,
\begin{equation}
G_{xy}=-\frac{e^2}{2\pi \hbar}\sum_{j}^{} C_j
\end{equation}
This is a well-established result\cite{PhysRevB.31.3372}.

Thus, the problem of the computation of the current can be reduced to the problem of the computation of the transmobility. It can be computed by solving the Schrodinger evolution equation for the wavefunction and computing the expectation value of the velocity operator that is defined as
\begin{equation}
\hat{\dot{x}}_\alpha=\frac{1}{\hbar}\frac{\partial \hat{H}}{\partial q_\alpha}
\end{equation}
where $q_\alpha$ is a quasi-momentum parameter of the system. In the case of the milti-terminal superconducting junction it would be the global phase of one of the leads\cite{ncomms11167}.

The computation of transmobility of a particle in a single band is standard. A weak force doesn't cause interband transitions but changes the wavevector in time $\hbar \dot{q}_\alpha=F_\alpha$. For constant $F_\alpha$
\begin{equation}
\hbar\vec{q}(t)=\vec{F} t+{\rm const}
\label{sweep}
\end{equation}
for a resulting trajectory in $\vec{q}$-space sweeps over the whole Brillouin zone at long time scale $t\sim \hbar \sqrt{\Omega}/|\vec{F}|$ at least for incommensurate direction of force.

In the case when the initial state is the eigenstate of the Hamiltonian, the mean current reduces to the Berry curvature of the initial state. It follows from the semi-classical equations of motion\cite{PhysRevB.59.14915} that the Berry curvature brings an addition to the velocity:
\begin{equation}
	\vec{\dot{x}}=\frac{1}{\hbar}\frac{\partial E(q,x)}{\partial \vec{q}}-B^{\alpha \beta} \dot{q}_\beta=\frac{1}{\hbar}\frac{\partial E(q,x)}{\partial \vec{q}}-B^{\alpha \beta} \frac{F_\beta}{\hbar}
	\label{dotx}
\end{equation}
We consider this equation at the long time scales such that the whole Brillouin zone is swept over. Upon the sweeping desribed by Eq.$\eqref{sweep}$ the first term in $\eqref{dotx}$ describes Bloch oscillations and averages to zero since it is a derivative over $\vec{q}$. The second term reduces to the average of the Berry curvature 
\begin{equation}
\frac{1}{\Omega}\int d\vec{q} B^{\alpha \beta}=\frac{2\pi C e^{\alpha \beta}}{\Omega}
\end{equation}
which is an integer Chern number $C$, and gives the drift in the direction perpendicular to the applied force
\begin{equation}
	\langle \dot{x}_\alpha\rangle=-\frac{2\pi C}{\hbar\Omega}\times \epsilon_{\alpha \beta}F^\beta,\quad \mu=-\frac{2\pi C}{\hbar \Omega}
\end{equation}
in agreement with Eq.$\eqref{mu}$.

Let us compute the transmobility for a superposition. One can make such a superposition by an oscillating modification\cite{rqm} of bandstructure parameters. A pulse of these oscillations brings a number of particles to the superposition state in a narrow region in $\vec{q}$-space where the frequency of the oscillations matches the energy difference between the bands $|\psi_{0,1}\rangle$ with different Chern numbers $\hbar\omega=\epsilon_1(\vec{q}_0)-\epsilon_0(\vec{q}_0)$. After the pulse, wavefunction of one particle with the quasi-momentum $\vec{q}_0$ is
\begin{equation}
|\psi(t=0)\rangle=a|\psi_0(\vec{q}_0)\rangle+b|\psi_1(\vec{q}_0)\rangle
\end{equation}
$a,b$ being the superposition coefficients. We need to solve
\begin{equation}
i\hbar \frac{\partial \psi(t)}{\partial t}=H(\vec{q}(t))\psi(t),\quad \vec{q}(t)=\vec{q}_0+\vec{F}t
\label{equation}
\end{equation}
where we treat the quasi-momentum of the particle as an adiabatically changing parameter of the Hamiltonian. We seek for the wavefunction in the instantaneous eigenbasis of $\hat{H}(t)$, $|\psi(t)\rangle=\sum_{k}^{}c_k(t)|\psi_k(t)\rangle$, $k$ labeling all the bands. The coefficients $c_k$ in the zero order of adiabatic perturbation theory in the parameters $\dot{q}_{\alpha,\beta}$ are
\begin{equation}
c_0^{(0)}(t)=e^{i\theta_0(t)}a,\quad c_1^{(0)}(t)=e^{i\theta_1(t)}b,\quad c_{l>1}^{(0)}(t)=0
\label{c0}
\end{equation}
where
\begin{equation}
\theta_k(t)=-\int_{}^{t}d\tau E_k(t)+\int^t \vec{A}^{(k)}\cdot d\vec{q}(t)
\end{equation}
incorporates both the dynamical and geometric phases\cite{Berry45}, and $\vec{q}(t)$ being the path in parameter space as in Eq.$\eqref{equation}$. Here $A_\alpha^{(k)}$ is the Berry connection, $A_\alpha^{(k)}=i\langle k|\partial_{q_\alpha} k\rangle $. To compute the expectation value of the drift velocity, $\langle \psi(t)|\hat{\dot{x}}_\alpha|\psi(t)\rangle$, we need to evaluate $c_k$ up to the first order 
\begin{equation}
c_0^{(1)}(t)=\frac{ibe^{i\theta_1(t)}\langle \psi_0|\dot{\psi}_1\rangle}{E_0-E_1},\quad c_1^{(1)}(t)=\frac{iae^{i\theta_0(t)}\langle \psi_1|\dot{\psi}_0\rangle}{E_1-E_0}
\label{c01}
\end{equation}
\begin{equation}
c_{l>1}^{(1)}(t)=\frac{ibe^{i\theta_1(t)}\langle \psi_l|\dot{\psi}_1\rangle}{E_l-E_1}+\frac{iae^{i\theta_0(t)}\langle \psi_l|\dot{\psi}_0\rangle}{E_l-E_0}
\label{cl}
\end{equation}
We average this expectation value over the short time scale $t_s|E_0-E_1|\gg \hbar$ thereby neglecting the fast oscillating terms $\sim e^{i(\theta_{0}(t)-\theta_{1}(t))}$. We obtain
\begin{equation}
\langle  \hat{\dot{x}}_\alpha \rangle=\frac{1}{\hbar}(|a|^2\frac{\partial E_0}{\partial q_\alpha}+|b|^2\frac{\partial E_1}{\partial {q_\alpha}}-F^\beta[|a|^2 B_{\alpha \beta}^{(0)}+|b|^2B_{\alpha \beta}^{(1)}])
\label{weited}
\end{equation}
where the first two terms are of zero order in $\dot{q}_{\alpha,\beta}$ and coming from Eq.$\eqref{c0}$ and the last two are the first order adiabatic correction coming from Eqs.$\eqref{c01},\eqref{cl}$. Thus, Eq.$\eqref{weited}$ generalizes Eq.$\eqref{dotx}$ for the case of a superposition. As above, we consider the drift velocity at the long time scale such that the whole Brillouin zone is swept over. Upon the sweeping described by Eq.$\eqref{sweep}$ the first two terms in $\eqref{weited}$ average to zero since they are derivatives over $\vec{q}$ with time-independent coefficients. The last two terms reduce to the weighted sum of the averages of two Berry curvatures. Finally, we obtain
\begin{equation}
\langle \dot{x}_\alpha\rangle=-\frac{2\pi C^\prime}{\hbar\Omega} \times \epsilon_{\alpha \beta}F^\beta,\quad C^\prime=|a|^2C_0+|b|^2 C_1
\end{equation}
Thus, the transmobility of a particle in a superposition state is not quantized. The absence of the topological quantization can be explained from the fact that the dynamical superposition is not periodic in $\vec{q}$ even if upon sweeping the wavefunction comes to the same or close point in Brillouin zone. It won't be the same due to the fast oscillating factors $e^{i\theta_{0,1}(t)}$. In other words, the dynamical superpositions do not form a manifold where topological constraints can be imposed.

The experimental observation of this effect is rather straightforward. If we apply sequence of pulses to a sample a number of particles will be brought to a superposition state. This number will be proportional to the intensity of radiation. One would just see deviations from the quantized value of the current proportional to the radiation intensity.

\section{Topological constraint on the mixing matrix element\label{Sec:propt}}
In the previous Section, we have seen that a dynamical superposition shows a non-topological response. Since it is not an eigenfunction of a Hamiltonian. In this Section we consider the static superpositions that are the eigenfunction of a stationary Hamiltonian. We consider a smooth $N\times N$ Hamiltonian $H_0+H^\prime$ with $N\ge 2$. The Hamiltonian $H_0$ is assumed to be diagonalized giving rise to a bandstructure that includes topologically non-trivial bands. The addition $H^\prime$ is a perturbation that is generally non-diagonal in this basis. With this, the eigen-bands of the total Hamiltonian will be quantum superpositions of the eigenbands of the unperturbed Hamiltonian with well-defined energies. As above, all the bands and the Hamiltonian are defined on 2-dimensional compact space of parameters $q_{1,2}$. Let us concentrate on two bands $|\psi_{0,1}(\vec{q})\rangle$ with different Chern numbers $C_{0,1}=0,1$ introduced above. 

We can choose different gauges for wavefunctions in these bands by multiplying it with a phase factor $\chi(\vec{q})$. By a proper choice of the gauge the topologically trivial band $|\psi_0(\vec{q})\rangle$ can be made not only quasi-continuous but truly continuous. In distinction from this the topologically non-trivial band $|\psi_1(\vec{q})\rangle$ cannot be made continuous everywhere. However, by a proper gauge choice it is possible to make it continuous within the Brillouin zone placing possible discontinuities on its boundary. We will stick to this convenient gauge choice. The effective Hamiltonian in the subspace of those two bands $H$ is obtained by projecting $H_0+H^\prime$ on that subspace
\begin{equation}
H=\begin{pmatrix}
\epsilon_0(\vec{q}) & t(\vec{q})\\
t^*(\vec{q}) & \epsilon_1(\vec{q})
\end{pmatrix}
\label{2x2}
\end{equation}
where the mixing matrix element
\begin{equation}
t(\vec{q})=\langle \psi_0|H^\prime(\vec{q})|\psi_1\rangle
\label{tmatrelem}
\end{equation}
is a continuous function inside the Brillouin zone by virtue of the gauge choice made. 

We will prove now a general and important topological constraint imposed on $t(\vec{q})$: if the Chern numbers of two bands are different, then for any $H^\prime$ there must exist a point in $\vec{q}$-space where the mixing matrix element vanishes
\begin{equation} 
\exists(q_1^*,q_2^*): t(q_1^*,q_2^*)=0
\end{equation}
For simplicity, let us consider the case when the parameter space is a torus corresponding to a Brillouin zone of a 2-dimensional crystal, generalization to other types of surfaces is straightforward. According to the general theory of characteristic classes, the wavefunction with a nontrivial Chern number has to have a singularity at some point in parameter space $\vec{q^\prime}$. By choosing the gauge described above we have moved the singularity of $|\psi_1(\vec{q})\rangle$ to the boundary of the 2-disc from which the torus is then obtained by gluing the sides. The wavefunction is continuous inside the Brillouin zone then but not periodic and the boundary conditions are given by
\begin{equation}
\psi(0,q_2)=e^{i\theta(q_2)}\psi(2\pi,q_2),\quad \psi(q_1,0)=e^{i\theta(q_1)}\psi(q_1,2\pi)
\end{equation}
the winding of the phase $\theta(q_1,q_2)$ along the boundary yields precisely the first Chern number. Then, according to $\eqref{tmatrelem}$ the mixing matrix element $t(q_1,q_2)$ also acquires the same phase winding along the boundary. Due to the discrete nature of this winding it does not change upon smooth variations of the parameters, so one can smoothly deform the contour on which the winding is defined. One will not be able to shrink this contour to a point if $t(q_1,q_2)\ne 0$ everywhere since due to the conservation of the winding. In this case one would obtain a discontinuity of $t(q_1,q_2)$ at some point unless $t(q_1,q_2)=0$ in this point. So, this proves the topological constraint discussed.
\section{Band crossing: topological transition and Berry curvature distribution\label{Sec:static}}
In this Section we consider the avoided crossing of two bands of different topology. We assume that the values of the mixing matrix elements in $\eqref{2x2}$ $t(\vec{q})$ are small in comparison with typical width of the bands $\epsilon_{0,1}(\vec{q})$. More precisely, 
\begin{equation}
|t(q_1,q_2)|\ll \max_{\vec{q}}\epsilon_{0,1}(\vec{q})- \min_{\vec{q}}\epsilon_{0,1}(\vec{q})
\end{equation}

We would like to move the energies of the bands with respect to each other. Let us assume that both energies depend on an additional parameter $\eta$ 
\begin{equation}
\epsilon_{0,1}(\vec{q})=\epsilon_{0,1}^0(\vec{q})\pm \eta
\end{equation}
As an example of a concrete physical situation where it can be realized one can consider a bilayer material with weak tunnel coupling between the layers. The bands $\epsilon_0$ and $\epsilon_1$ are situated in different layers, weak tunneling is responsible for matrix mixing elements and to $2\eta$ corresponds to the difference of electrostatic potentials between the layers that can be induced by a perpendicular electric field.

We see that at sufficiently large $|\eta|$ the unperturbed band energies never cross: $\eta \to -\infty$ $\epsilon_0(q_1,q_2)<\epsilon_1(q_1,q_2)$ for all $(q_1,q_2)$, for $\eta \to +\infty$ $\epsilon_0(q_1,q_2)>\epsilon_1(q_1,q_2)$ for all $(q_1,q_2)$. By changing $\eta$ from large negative to large positive values we move the energies of the bands with respect to each other and make them cross in a certain interval of $\eta$. 

Let us consider how the Chern numbers of the eigenbands of $\eqref{2x2}$ change upon changing $\eta$. When the energies of the unperturbed bands do not cross at any point ($|\eta|\to +\infty$) the mixing can be neglected and the Chern numbers of the eigenbands the same as without mixing. However, we see that the topological configurations of the bands are different for $\eta \to \pm \infty$. At $\eta\to -\infty$ the topological charge is concentrated in the lower band while it is transfered to the upper band when $\eta \to +\infty$. We conclude that the topological transition must occur upon the band crossing. 

\begin{figure}
	\centerline{\includegraphics[width=0.45\textwidth]{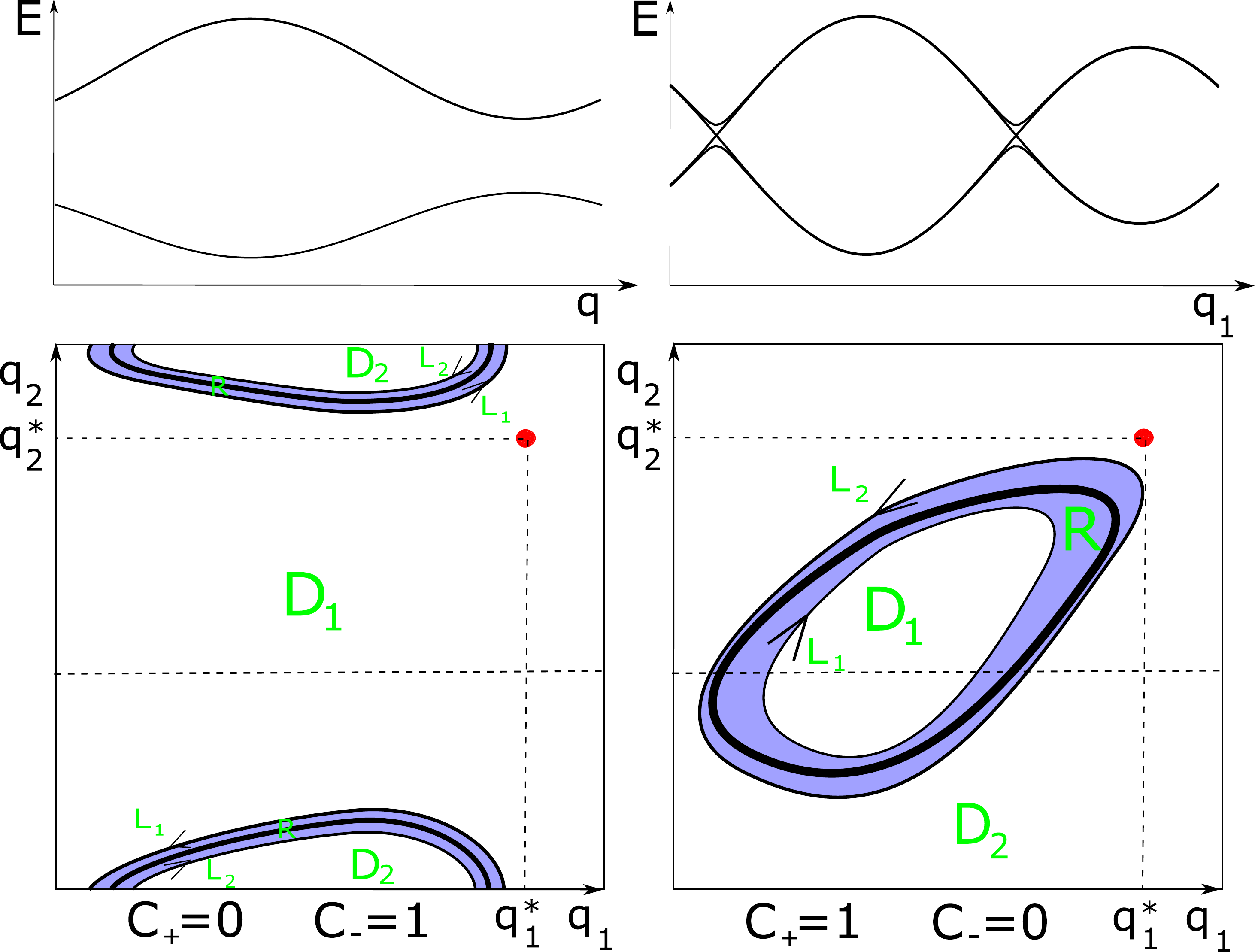}
	}
	\caption{Right: separation of the $(q_1,q_2)$ parameter space into 3 regions before the topological transition: $D_1$ where $\epsilon_1<\epsilon_0$, $D_2$ where $\epsilon_1>\epsilon_0$ and a narrow strip $R$ around the line $\epsilon_1=\epsilon_0$. The width of the strip is determined by the value of mixing $t$ in $\eqref{2x2}$. Oriented boundaries of $D_{1,2}$ are denoted as $L_{1,2}$. Left: after the transition}
	\label{areas}
\end{figure}

The crossing of the bands is generally avoided at a given value of $\eta$. The unperturbed bands cross at the lines where $\epsilon_0(q_1,q_2)=\epsilon_1(q_1,q_2)$, and the area of the space of $(q_1,q_2)$ is separated into parts by these lines. Except for the close vicinity of these lines, the wavefunctions are expected to be close to with the unperturbed ones since the mixing is small. At the lines the crossings are generally avoided and the bands do not actually cross (see Fig.\ref{areas}) being separated by an energy difference at least $2|t(\vec{q})|$. The bandmixing is strong in a narrow strip that includes the lines. The typical width of the strip can be estimated as $|\delta \vec{q}_M|\simeq |\frac{t(\vec{q})}{\partial(\epsilon_1(\vec{q})-\epsilon_0(\vec{q}))/\partial q_\alpha}|$. The Chern numbers can be ascribed to the resulting lower and upper energy bands and do not change while the crossing is avoided. 

There is however a critical value of $\eta=\eta_c$ at which the crossing lines intersect the special point $\vec{q}^*$ mentioned in the previous Section. At this point $|t(\vec{q}^*)|=0$ and the crossing is not avoided. Two bands are connected at the point $\vec{q}^*$ and become a single band with the topological charge 1. This is the point of topological phase transition. 

In the rest of the Section we consider the distribution of the Berry curvature in the $\vec{q}$-space for the situation of the band crossing. The Chern number is proportional to the integral of the Berry curvature over the space. A naive consideration would neglect the small mixing, so the Chern number of the eigenband reduces to the sum of integrals of Berry curvatures of the unperturbed bands over the corresponding regions. However, this sum is by no means integer. 

So motivated, let us consider the situation in detail. Let us assume that the condition for the crossing of the unperturbed levels $\epsilon_0=\epsilon_1$ is satisfied along a single closed line in the $(q_1,q_2)$ space (see Fig.\ref{areas}). Then, under the assumptions described above away from this line the wavefunctions should approach the non-mixed functions up to a phase factor $|\psi_\pm(q_1,q_2)\rangle\to e^{i\chi_{0,1}(q_1,q_2)}|\psi_{0,1}(q_1,q_2)\rangle$. We denote these areas as $D_{1,2}$. We denote the values of the Berry curvature integrals over these areas as
\begin{equation}
I_i^j=\frac{1}{2\pi}\int_{D_i} B^{(j)}_{12}dq_1dq_2
\end{equation}
$j=0,1$ labeling the unperturbed bands. If we neglect a narrow area that encloses the energy crossing line (denoted as $R$ in Fig.\ref{areas}). 
the "Chern numbers" of the upper ($\tilde{C}_+$) and lower ($\tilde{C}_-$) bands do not reduce to integers
\begin{equation}
\tilde{C}_+=I_1^{(0)}+I_2^{(1)}\ne C_1=1=I_1^{(1)}+I_2^{(1)}
\end{equation}
\begin{equation}
\tilde{C}_-=I_1^{(1)}+I_2^{(0)} \ne C_0=0=I_1^{(0)}+I_2^{(0)}
\end{equation}
in general, which seems to contradict the general topology statement. 

This paradox is resolved by considering the Berry curvature in region $R$. There, the bands are strongly mixed and thus this narrow region brings a finite contribution to the Chern number so that the typical Berry curvature in this region is $\sim |\delta \vec{q}_M|^{-1}$, that is much bigger that the typical Berry curvature in the regions $D_{1,2}$. We denote the contributions from the region $R$ as
\begin{equation}
I^R_\pm=\frac{1}{2\pi}\int_R B^\pm_{12}dq_1dq_2
\end{equation}
The true topological charges $C_\pm$ of the eigenbands $|\psi_\pm\rangle$ are
\begin{equation}
C_+=I_1^{(0)}+I_2^{(1)}+I^R_+
\label{cplus}
\end{equation}
\begin{equation}
C_-=I_1^{(1)}+I_2^{(0)}+I^R_-
\label{cminus}
\end{equation}
and must be integer. 

To see that we shall evaluate $I^R_\pm$. In order to do this, we consider the Berry connections of the superpositions. It is defined as $A^\pm_\alpha=i\langle \psi_\pm|\partial_{q_\alpha} \psi_\pm \rangle$. For two superpositions under consideration, $|\psi_\pm \rangle=a_\pm|\psi_0\rangle+b_\pm|\psi_1\rangle$ where superposition coefficients $a_\pm,b_\pm$ are obtained by diagonalizing $\eqref{2x2}$. We express the connections in terms of superposition coefficients
\begin{equation}
A^\pm_\alpha=ia_\pm^*\partial_{q_\alpha} a_\pm+ib_\pm^*\partial_{q_\alpha} b_\pm+|a_\pm|^2 A_\alpha^{(0)}+|b_\pm|^2 A_\alpha^{(1)}+
\notag
\end{equation}
\begin{equation}
+(i a_\pm^* b_\pm\langle \psi_0|\partial_{q_\alpha} \psi_1\rangle+c.c)
\label{asuperp}
\end{equation}
We see that in addition to the weighted sum of connections there are extra contributions to $A^\pm_\alpha$. We will see that they are responsible for the resulting Chern numbers being integer. At distances from the crossing line that $\gg |\delta \vec{q}_M|$ but still much smaller than the typical size of the Brillouin zone, the coefficients $a_-$ and $b_+$ vanish in $D_1$ and the coefficients $a_+$ and $b_-$ vanish in $D_2$ (see Fig.\ref{areas}). With this we obtain the following asymptotics for $\vec{q} \in D_1$
\begin{equation}
\begin{cases}
&A^+_\alpha=i\partial_{q_\alpha} (\log a_+)+A_\alpha^{(0)} \\
&A^-_\alpha=i\partial_{q_\alpha} (\log b_-)+A_\alpha^{(1)}
\end{cases}
\end{equation} 
and for $\vec{q} \in D_2$
\begin{equation}
\begin{cases}
&A^+_\alpha=i\partial_{q_\alpha} (\log b_+)+A_\alpha^{(1)} \\
&A^-_\alpha=i\partial_{q_\alpha} (\log a_-)+A_\alpha^{(0)}
\end{cases}
\label{apm}
\end{equation} 
We denote the contour integrals of the Berry connections over two oriented boundaries $L_i=-\partial D_i$ of the regions $D_i$
\begin{equation}
J_i^j=-\frac{1}{2\pi}\int_{L_i}^{}\vec{A}^{j}\cdot d\vec{q}
\label{contints}
\end{equation}
It is crucial to note that in the absence of singularities of all the functions away from the boundary of the ($q_1,q_2$)-space one may use the Stokes theorem to reduce the surface integrals of the Berry curvature $I_i^j$ to the contour integrals of the Berry connection. Applying this we express the contributions of $D_i$ in terms of the contour integrals $\eqref{contints}$
\begin{equation}
\begin{cases}
&I_1^j=J_1^j\notag \\
&I_2^j=J_2^j+C_j
\end{cases}
\end{equation}
As for the contributions from region $R$, they are expressed in terms of $I_i^j$ and the contour integrals of the gradients of the phases of superposition coefficients (the latter holds since $|a_-|,|b_+|\to 1$ in $D_2$ and $|a_+|,|b_-|\to 1$ in $D_1$)
\begin{equation}
I^R_+=\frac{1}{2\pi}\int_{L_1}\vec{A}^+\cdot d\vec{q}+\frac{1}{2\pi}\int_{L_2}\vec{A}^+\cdot d\vec{q}=
\notag
\end{equation}
\begin{equation}
=-I_2^{(1)}+C_1-I_1^{(0)}+\frac{i}{2\pi}\int_{L_2}^{}d\vec{l}\cdot\vec{\nabla} \log b_++
\notag
\end{equation}
\begin{equation}
+\frac{i}{2\pi}\int_{L_1}d\vec{l}\cdot\vec{\nabla} \log a_+
\label{I52}
\end{equation}
\begin{equation}
I^R_-=\frac{1}{2\pi}\int_{L_1}\vec{A}^-\cdot d\vec{q}+\frac{1}{2\pi}\int_{L_2}\vec{A}^-\cdot d\vec{q}=
\notag
\end{equation}
\begin{equation}
=-I_2^{(0)}+C_0-I_1^{(1)}+\frac{i}{2\pi}\int_{L_2}^{}d\vec{l}\cdot\vec{\nabla} \log a_-+
\end{equation}
\begin{equation}
+\frac{i}{2\pi}\int_{L_1}d\vec{l}\cdot\vec{\nabla} \log b_-
\label{I51}
\end{equation}
The last two terms in Eqs. $\eqref{I51}$ and $\eqref{I52}$ are integer multiples of $2\pi$ since they are equal to the sum of the windings of the phases of superposition coefficients along the closed contours $L_{1,2}$. In order to compute these phases, we diagonalize the effective Hamiltonian in the vicinity of the crossing line
\begin{equation}
H^{({\rm eff})}=\begin{pmatrix}
\varepsilon & t\\
t^* & -\varepsilon
\end{pmatrix}
\label{heff}
\end{equation}
where the small energy difference is $\varepsilon=(\epsilon_0-\epsilon_1)/2$, the limits $\varepsilon/|t|\to \pm\infty$ bring us to the region $D_{1}$, $D_2$ correspondingly. We approximate the mixing matrix element by its value exactly at line disregarding its dependence on $\varepsilon$. We diagonalize $\eqref{heff}$ to find two eigenfunctions $|\psi_\pm \rangle=a_\pm(\varepsilon)|\psi_0\rangle+b_\pm(\varepsilon)|\psi_1\rangle$. To remove the ambiguity of the phases we will choose the phase of $|\psi_+\rangle$ to coincide with the phase of $|\psi_1\rangle$ and the phase of $|\psi_-\rangle$ to coincide with the phase of $|\psi_0\rangle$ in the region $D_2$ so that
\begin{equation}
a_-(-\infty)=1,\quad b_+(-\infty)=1
\label{apm2}
\end{equation}
Diagonalizing $\eqref{heff}$ with the phase fixing conditions $\eqref{apm2}$, we obtain the superposition coefficients in the region $R$
\begin{equation}
\begin{cases}
b_-(\varepsilon)=\notag\\
a_-(\varepsilon)=
\notag
\end{cases}\frac{1}{{\sqrt{|t|^2+(\varepsilon+\sqrt{|t|^2+\varepsilon^2})^2}}}
\begin{array}{c}
\frac{-t^*(\varepsilon+\sqrt{\varepsilon^2+|t|^2})}{|t|}\notag\\
|t|
\end{array}
\end{equation}
\begin{equation}
\begin{cases}
b_+(\varepsilon)=\notag\\
a_+(\varepsilon)=
\notag
\end{cases}\frac{1}{{\sqrt{|t|^2+(-\varepsilon+\sqrt{|t|^2+\varepsilon^2})^2}}}
\begin{array}{c}
\sqrt{|t|^2+\varepsilon^2}-\varepsilon\notag\\
t
\end{array}
\end{equation}
With this we can find their asymptotics at $\varepsilon\to +\infty$ that correspond to the values of superposition coefficients on $L_1$
\begin{equation}
b_- \to -t^*/|t|,\quad a_+\to t/|t|
\label{apm1}
\end{equation}
with this one can compute the last contributions in $\eqref{I51}$ and $\eqref{I52}$
\begin{equation}
\int_{L_2}^{}d\vec{l}\cdot\vec{\nabla} \log a_-=0=\int_{L_2}^{}d\vec{l}\cdot\vec{\nabla} \log b_+
\label{w1}
\end{equation}
\begin{equation}
\frac{i}{2\pi}\int_{L_1}^{}d\vec{l}\cdot\vec{\nabla} \log a_+=-\frac{i}{2\pi}\int_{L_1}^{}d\vec{l}\cdot\vec{\nabla} \log b_-=
\notag
\end{equation}
\begin{equation}
=\int_{L_1}^{}d\vec{l}\cdot\vec{\nabla} \log t=w
\label{w2}
\end{equation}
where we denoted the winding of the phase of the mixing matrix element along $L_1$ as $w$. Before the transition when the point $\vec{q}^*$ is outside $L_1$ we have $w=0$. After the transition by definition of the mixing matrix element $\eqref{tmatrelem}$ the value of $w$ coincides with the value of $C_1$ with the opposite sign. Then from Eqs. $\eqref{I51}$, $\eqref{I52}$, $\eqref{w1}$, $\eqref{w2}$ we obtain
\begin{equation}
C_+=1+\frac{w}{2\pi},\quad C_-=-\frac{w}{2\pi}
\label{cpm}
\end{equation}
So, in the case before the transition we have 
\begin{equation}
C_+=1,\quad C_-=0
\end{equation}
and after the transition
\begin{equation}
C_+=0,\quad C_-=1
\end{equation}
as shown in Fig.$\ref{areas}$. The above reasonings can be straightforwardly generalized to the crossing of two bands with arbitrary Chern numbers $m,n$. We note that in this case the mixing matrix element has $|n-m|$ zeroes, so we expect $|n-m|$ phase transitions upon the crossing.
\section{Many bands: general properties of the phase diagrams\label{Sec:moar}}
In this Section, we consider the topological phases and the transitions between the phases for an arbitrary number of bands $N$. Let us assume that the bandstructure depends on $M$ additional parameters. 
At arbitrary point in the space of $M$ parameters the bands do not cross and can be sorted in the order of increasing energy. The first Chern numbers of each band are well-defined. With this, we define a topological phase as an enumeration of the first Chern numbers of the bands in the order of ascending energy. It is convenient to describe it with a Greek letter milti-index that consists of enumeration of the first Chern numbers of the bands in the order of ascending energy (e.g. $\beta\equiv \{n_i\}$, $i=1,..,N$). 

To achieve a 2-band generic crossing one has to tune 3 independent parameters\cite{theorem}, two of them might be quasi-momenta. Therefore, the band crossings occur in a ($M-1$)-subspace of the space of additional parameters and this is a subspace of the topological transition points. 

For many bands there can be the singularities of higher order, e.g. 3- and 4-bands crossing. They occur in the subspaces of dimension $M-6$ and $M-13$. We restrict our consideration to smaller dimensions where all transitions correspond to pairwise crossings of the bands. 

\begin{figure}
	\centerline{\includegraphics[width=0.45\textwidth]{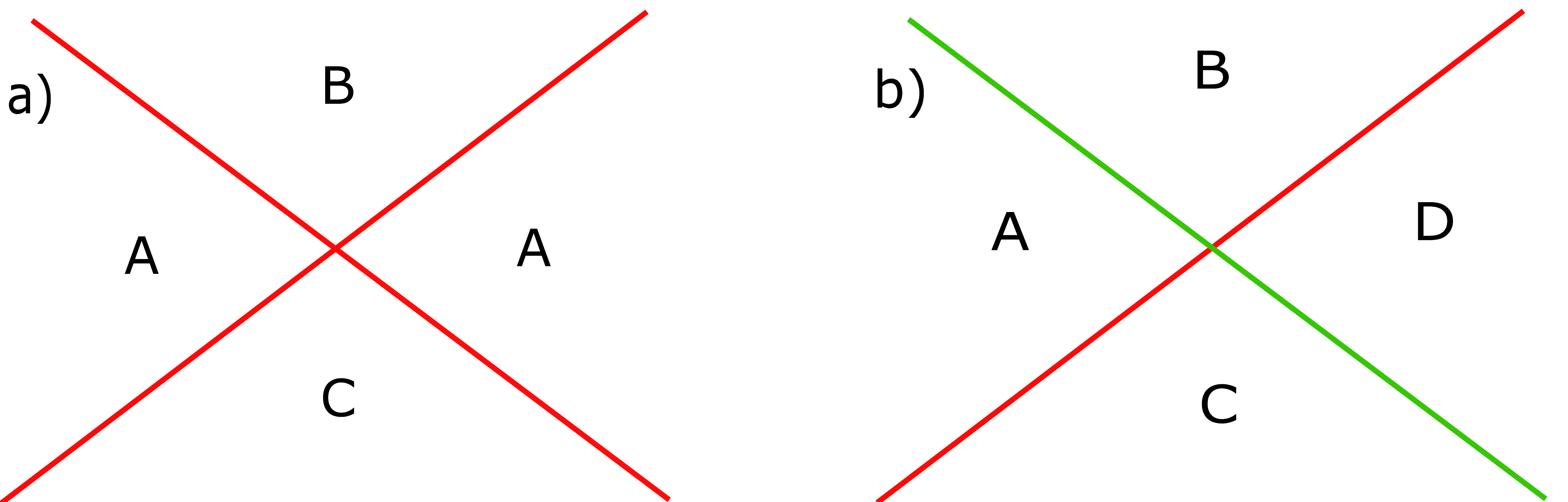}
	}
	\caption{Triple points are generically absent in the phase diagrams under consideration. The critical points are quadruple with 4 phases meeting at the point. There are two types of the quadruple points. a) Type-I: crossing of the lines of the same color. always connects three different phases. The crossing point is stable upon small variations of parameters. The exchange of Chern number by 1 guarantees that two of the four phases are the same and the only possibility for phases is: $A=(...,n_i,n_{i+1},...)$, $B=(...,n_i\pm 1,n_{n+1}\mp 1,...)$ and $C=(...,n_i\mp 1, n_{i+1}\pm 1,...)$. b) Type-II: crossing of the lines of different colors. All 4 phases are different. There may be several options for those phases depending on colors and signs of Weyl points, e.g.: $A=(...,n_i,...,n_j,...)$, $B=(...,n_i-1,n_{i+1}+1,...,n_j,...)$, $C=(...,n_i,...,n_j-1,n_{j+1}+1,...)$ and $D=(...,n_i-1,n_{i+1}+1,...,n_j-1,n_{j+1}+1)$}
	\label{crosses}
\end{figure}%

We start with $M=1$. Two quasi-momenta and single additional parameter form a 3-dimensional parameter space. The 2-band crossings correspond to isolated Weyl points\cite{weyl} in this 3-dimensional space. These points are topologically stable bearing a topological charge related to the point-like divergence of the 3-dimensional Berry curvature. Two bands exchange Chern number 1 upon the change of the additional parameter. Let us note that the crossings of different pairs of the bands bear distinct topological charges. We will call them colors: there are $N-1$ distinct colors. Generically the positions of Weyl points do not coincide. The accidental coincidence of the positions of two Weyl points corresponding to two pairs with one mutual band would lead to a 3-band crossing at this mere point. As discussed, that we do not consider. The phase diagram consists of intervals separated by the projections of the Weyl poins onto the axis of additional parameter. This implies the following rule: the phase transition cannot occur between two arbitrary phases, since the transitions involve the Chern number exchange between neighboring bands ($\{...,n_i,n_{i+1},...\}\leftrightarrow \{...,n_i+1,n_{i+1}-1,...\}$)

Let us consider $M=2$ and phase diagrams in the space of two additional parameters $s_{1,2}$. Weyl points develop into singularity lines in the resulting 4-dimensional space. These lines come in distinct colors. The phase diagram is obtained by the projection of these lines onto $(s_1,s_2)$ plane. Therefore, the critical points where more than two phases coexist, are quadruple (see Fig.$\ref{crosses}$). In such a critical point there are two 2-band crossings in the Brillouin zone. Generically these crossings occur at different $\vec{q}$. 

There are two types of critical points. Type-I corresponds to the crossing of two lines with the same color (Fig. $\ref{crosses}$(a)). Since each line corresponds to the exchange of Chern numbers by 1 in the neighbouring bands, 2 of the 4 phases meeting at the point must be the same. For instance, two identical phases $A=(...,n_i,n_{i+1},...)$ and two different ones: $B=(...,n_i\pm 1,n_{n+1}\mp 1,...)$ and $C=(...,n_i\mp 1, n_{i+1}\pm 1,...)$ (see Fig.$\ref{crosses}$). If we consider $M>2$ and give small variations to extra parameters ($s_3$, etc), these point will be stable with respect to these small variations of extra parameters. However, upon the larger variation of at least two extra parameters one can annihilate two 2-band crossings and thereby eliminate the critical point. Type-II corresponds to the crossings of the lines of distinct colors. In this case all 4 phases must be distinct. These points are even more stable than Type-I points. The Type-II crossing may change upon changing 4 extra parameters. More are needed to change the crossings of the lines of more distinct colors.

The most common features of the usual phase diagrams (e.g. for not topological phase transitions\cite{vol5}) are very distinct from the ones under consideration. In that case the critical points in 2-dimensional parametric space are triple.

\section{Example: bilayer Haldane model\label{Sec:haldane}}
To illustrate the above general considerations, we investigate the phase diagrams of a bilayer Haldane model. The Haldane model\cite{PhysRevLett.61.2015} for a single layer describes electrons in a periodic hexagonal lattice with two orbitals per site, ${\bf a}_i$ $i=1,2,3$ being the nearest neighbor vector distances and ${\bf b}_i$ being the next to nearest neighbor vector distances in the lattice. The Hamiltonian is a $\vec{q}$-dependent $2 \times 2$ matrix, the matrix structure describing 2 sublattices in hexagonal lattice, and reads
\begin{align}
H_H=
&\left\{ M+2t_{nnn}\sum_{j=1}^{3}\sin \vec{q}\cdot \vec{b}_j\right\}\tau_z+\notag\\
&+t_n\left\{\sum_{i=1}^{3}\cos \vec{q}\cdot \vec{a}_i\tau_x-\sin \vec{q}\cdot\vec{a}_i\tau_y\right\}
\label{2by2}
\end{align}
$M$ being the parameter corresponding to a mass coefficient, $t_n^{(\eta)}$ is the real amplitude of the next-neighbor hopping and a purely imaginary next-nearest neighbors hopping amplitude $t_{nnn}^{(\eta)}$. The Hamiltonian gives rise to two bands that are topologically trivial provided $|t_{nnn}|<|M/(3\sqrt{3})|$ and the lower and upper bands have Chern numbers $-{\rm sgn}(t_{nnn})$ and $+{\rm sgn}(t_{nnn})$ correspondingly for\cite{PhysRevLett.61.2015} $|t_{nnn}|>|M/(3\sqrt{3})|$. The band crossings occur at $t_{nnn}=\pm M/(3\sqrt{3})$ occur at high symmetry at the boundary of the Brillouin zone where $C_3$ rotational symmetry is preserved.

The phase diagram for the single layer model was investigated previously\cite{PhysRevLett.61.2015}. The phase diagram possessed two critical points of Type-I. According to the general analysis done in Sec$\ref{Sec:moar}$ those points are stable upon small variations of parameters of the model.

Stacking two layers together, arranging an energy shift and a tunnel coupling between the two will give rise to (avoided) band crossings and the associated topological transitions that we investigate. We put the layers exactly on the top of each other, exactly matching the site positions in lateral directions. This preserves the original $C_3$ rotational symmetry. We only take into account the tunneling between layers for the nearest neighbors. The Hamiltonian of the bilayer model in use is thus a $4\times 4$ matrix
\begin{equation}
\hat{H}=\begin{pmatrix}
H_H^{(1)} & {\rm T}\\
{\rm T}^\dagger & H_H^{(2)}
\end{pmatrix}
\label{4by4}
\end{equation}
the block structure is in the space of 2 layers. Each diagonal block is a single layer Haldane Hamiltonian with $M^{(\eta)}, t_n^{(\eta)}, t_{nnn}^{(\eta)}$ ($\eta=1,2$) and energy shifts $(\eta-1)\times s$.

The quasi-momentum-independent tunneling operator is diagonal in the sublattice space
\begin{equation}
{\rm T}=\begin{pmatrix}
T_a & 0\\
0 & T_b
\end{pmatrix}
\label{tunnel}
\end{equation}
We concentrate on the case ${\rm arg}(T_a)\ne {\rm arg}(T_b)$. If the phases of these two tunneling amplitudes are the same, the model possesses an extra degeneracy and does not suit to illustrate the generic situation. The degenerate case is addressed in Sec.$\ref{Sec:App1}$.

This Hamiltonian describes $N=4$ bands. The bandstructure depends on $M=10$ additional parameters: pairs of parameters describing the layers $t_n^{(\eta)}$, $t_{nnn}^{(\eta)}$ and $M^{(\eta)}$, absolute values of $T_{a,b}$ and their mutual phase difference and the energy shift $s$. 
For phase diagrams we need to choose two independent parameters. A natural parameter is the energy shift $s$ and the second natural parameter would be the bandwidth $W$. With this
\begin{equation}
t_n^{(\eta)}/a^{(\eta)}=t^{(\eta)}_{nnn}/b^{(\eta)}=M^{(\eta)}/c^{(\eta)}=W
\end{equation}
where we fix the coefficients $a^{(\eta)},b^{(\eta)},c^{(\eta)}$ to some values $\sim 1$. For some diagrams we implement an alternative choice where $M^{(\eta)}$ are fixed. 

One has to find the phase separation lines. To this end, one has to find the positions of the band crossings in a 4-dimensional parameter space $q_{1,2}$, $s$ and $W$. For the Hamiltonian $\eqref{4by4}$, this looks like a challenging numerical task. In fact, it is not. Owing to $C_3$ symmetry, a band crossing in an arbitrary point of the Brillouin zone would come in a triple. The associated phase transition would correspond to exchange of Chern numbers of 3 topological charges between the bands. This is an interesting possibility we have searched for yet didn't find it for the model under consideration. A possible reason for that is a general difficulty to achieve high topological numbers. In fact, in our examples we didn't see Chern numbers bigger than 1. 

Another possibility is to have the band crossings in the high-symmetry points $K$ or $K^\prime$. The associated topological transitions correspond to the exchange of unity topological charge, as in generic case. In these points, the eigenenergies can be readily found analytically,
\begin{align}
\epsilon_{1,2}=\frac{s+S}{2}\pm \sqrt{|T_a|^2+\left(\frac{s+A}{2}\right)^2}
\notag
\end{align}
\begin{align}
\epsilon_{3;4}=\frac{s-S}{2}\pm \sqrt{|T_b|^2+\left(\frac{s-A}{2}\right)^2}
\end{align}
where we define
\begin{equation}
S=M^{(1)}+M^{(2)}-3\sqrt{3}\sigma(t_{nnn}^{(1)}+t_{nnn}^{(2)})
\end{equation}
\begin{equation}
A=M^{(2)}-M^{(1)}-3\sqrt{3}\sigma(t_{nnn}^{(2)}-t_{nnn}^{(1)})
\end{equation}
$\sigma=\pm$ corresponding to $K$ and $K^\prime$, respectively. In this model the transition lines come in three colors. We associate the blue color with the crossing of the two lowest in energy bands, green color with that of the second and third band and the red color with that of the third and forth. The red lines emerge at
\begin{equation}
\epsilon_1=\epsilon_3
\end{equation}
for the blue lines emerge at
\begin{equation}
\epsilon_2=\epsilon_4
\end{equation}
and the green ones emerge either emerge at $\epsilon_1=\epsilon_4$ or $\epsilon_2=\epsilon_3$. At a given $W$ the 4 conditions above can be regarded as equations for $s$. Only 2 of 4 equations can have roots at either $K$ or $K^\prime$. This implies that at a given $W$ one finds $0$ or $2$ or $4$ phase transitions at different values of $s$.

\begin{figure}
	\centerline{\includegraphics[width=0.48\textwidth]{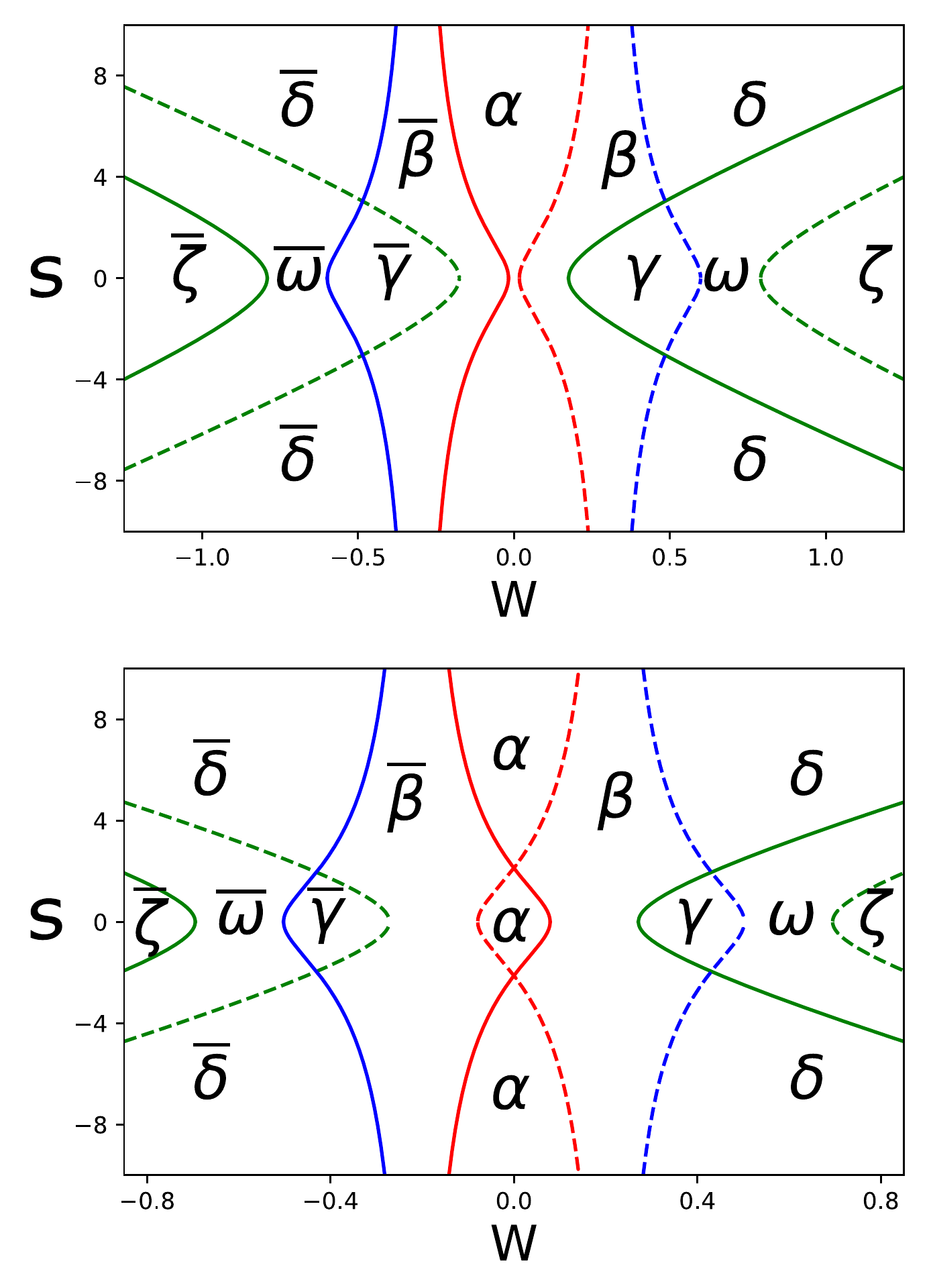}
	}
	\caption{Examples of the phase diagrams for the bilayer Haldane model. The phase transition lines of three colors (see the text) correspond to the crossings at $K$ point (dashed) or $K^\prime$ (solid). The parameters are: $b^{(1)}=b^{(2)}=0.5$, $T_a=0.5$, $T_b=2+0.2i$. Different choices of masses are made for the upper and lower panel, upper: $M^{(1)}=M^{(2)}=0.8$ and lower: $M^{(1)}=M^{(2)}=0.55$. The phases are given by enumeration of Chern numbers: $\alpha=(0,0,0,0)$, $\beta=(0,0,-1,1)$, $\gamma=(0,-1,0,1)$, $\delta=(-1,1,-1,1)$, $\zeta=(-1,-1,1,1)$, $\omega=(-1,0,0,1)$. Overline indicates the change of sign of all Chern numbers, e.g $\overline{\delta}=(1,-1,1,-1)$. The extra region of the phase $\alpha$ between two critical points seen in the lower panel disappears upon variation of the masses.}
	\label{fig2}
\end{figure}%
\begin{figure}
	\centerline{\includegraphics[width=0.48\textwidth]{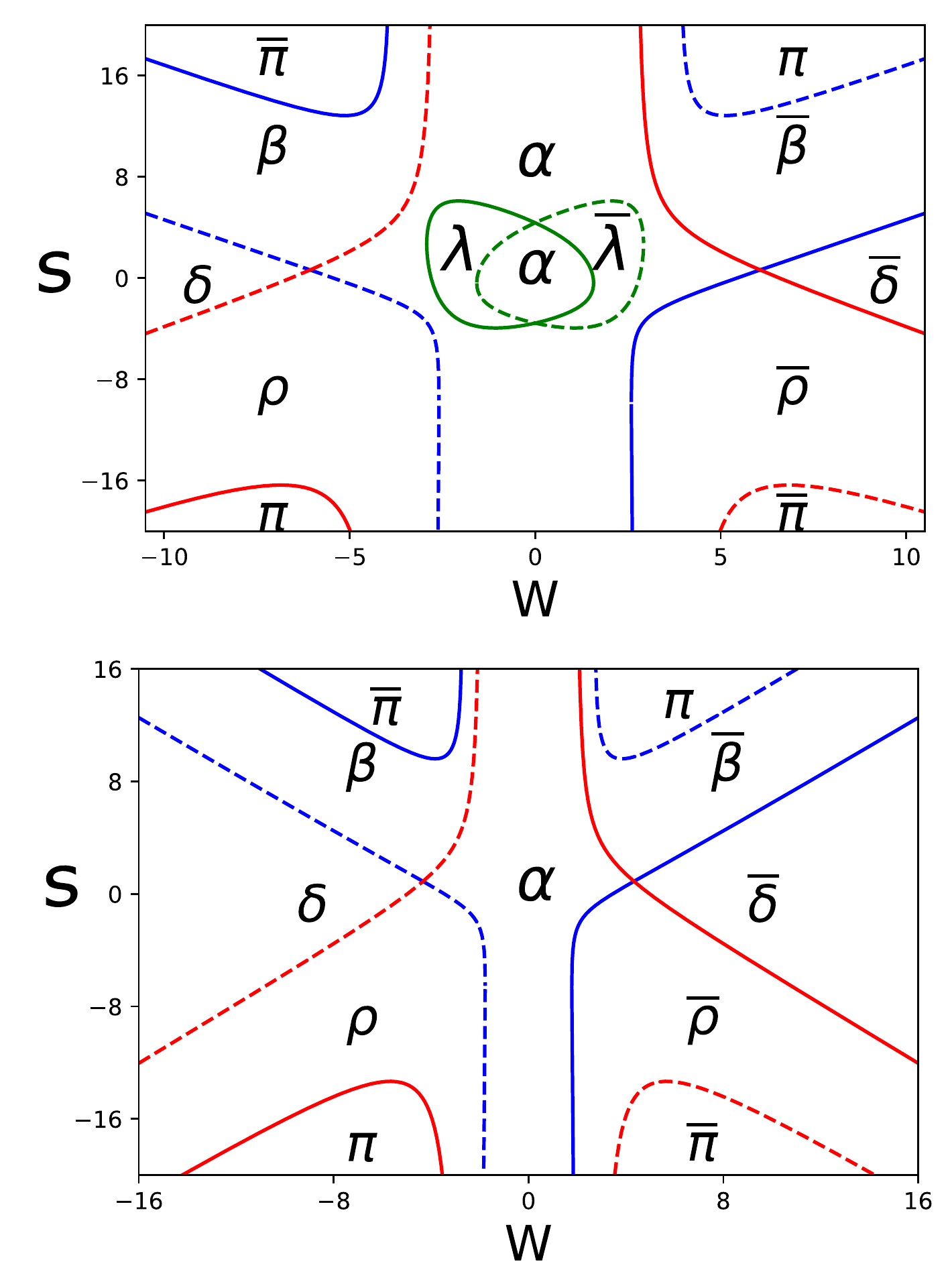}
	}
	\caption{Examples of the phase diagrams for the bilayer Haldane model. Choice of parameters $-3b^{(1)}=b^{(2)}=-0.3$, $T_a=3$, $T_b=1.5+0.2i$ and in the upper panel: $2M^{(1)}=M^{(2)}=4.2$ and in the lower panel: $2M^{(1)}=M^{(2)}=3$. In addition to the phases in Fig.$\ref{fig2}$ we also have: $\rho=(-1,1,0,0)$, $\pi=(-1,1,1,-1)$, $\lambda=(0,-1,1,0)$. The regions of the phases $\lambda,\overline{\lambda},\alpha$ in the center of the Figure disappears upon the variation of the masses.}
	\label{fig4}
\end{figure}%
\begin{figure}
	\centerline{\includegraphics[width=0.48\textwidth]{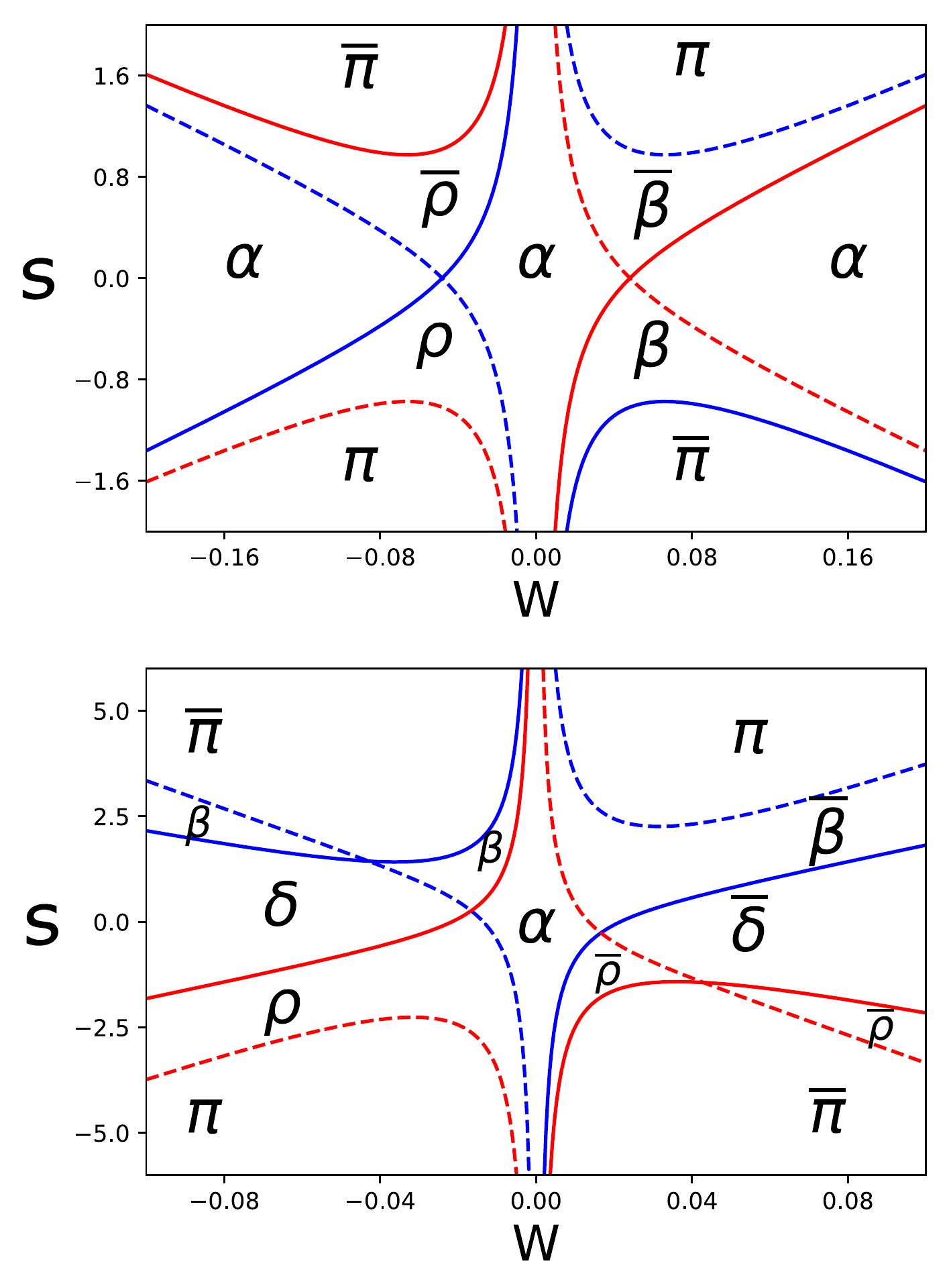}
	}
	\caption{Examples of the phase diagrams for the bilayer Haldane model. Choice of the parameters: $M^{(\eta)}=c^{(\eta)}*W$, $c^{(1)}=c^{(2)}=3.6$, $T_a=1+0.3i$, $T_b=0.3+1.5i$ and for upper panel $b^{(1)}=-b^{(2)}=5$, for lower panel $b^{(2)}=-10$. In the upper panel only the Type-I critical points are seen at $s=0$ and $W\approx \pm 0.05$. Upon variation of $b^{(2)}$ the blue and red solid lines interchange at positive as well as at negative $W$. This leads to the emergence of the Type-II critical points where the lines of different color cross and the non-compact regions of the phases $\delta$ and $\overline{\delta}$.}
	\label{fig6}
\end{figure}

The examples of the phase diagrams are presented in Figs. $\ref{fig2}$, $\ref{fig4}$, $\ref{fig6}$. In all these diagrams we see the features predicted in Sec.$\ref{Sec:moar}$. The critical points are all quadruple. There are Type-I critical points where 3 phases coexist and Type-II critical points where all 4 phases are different. 

In the upper panel of Fig. $\ref{fig2}$ we see two red transition lines separating single domain of phase $\alpha$ from the phases $\beta$ and $\overline{\beta}$ correspondingly. Upon tuning a parameter continuously, we can make these 2 lines intersect. Two Type-I critical points are formed at $s=\approx \pm 2$ and $W=0$ in the lower panel of Fig.$\ref{fig2}$. We also see three domains of phase $\alpha$ instead of one as in the upper panel. 

With $\ref{fig4}$ we illustrate more complex scenarios of this kind. In the center of the top panel we see isolated compact domains of the phases $\lambda,\overline{\lambda}$ coexisting with the phase $\alpha$. Upon tuning a parameter continuously, we can make these domains disappear. In the lower panel these domains are absent. We stress that in this case two compact transition lines disappear together with a pair of Type-I critical points.

We illustrate another scenario with Fig. $\ref{fig6}$. There are two main distinctions between upper and lower panel. The first one is the presence of non-compact domains of phases $\delta,\overline{\delta}$ in the lower panel contrary to the case of upper panel. This is in contrast to the case of the Fig.$\ref{fig4}$ where the disappearing domains were compact. The second distinction between panels is the presence of the Type-II critical points in the lower one in addition to the Type-I critical points in the upper one. Both changes are produced upon continuous variation of the parameter of the model. We stress that in this case the Type-II crossings are not produced at finite values of parameters but come from infinitely large positive and negative $W$. This is in contrast to the case of Fig.$\ref{fig2}$ where the critical points were produced pairwise at a given point on the phase diagram.

\section{Summary and Conclusions \label{Sec:Sum}}
In this Article, we address the topological properties of superpositions of the quantum bands. This involves definition and values of the topological numbers of the superposition of bands.

The naive expectation for the Chern number of the superpositions of the states to also be a weighted sum of Chern numbers should fail due to the general theory of characteristic classes. Therefore, this is a problem of general interest and we investigate the topological properties of the superpositions created in different ways.

The first way is to create the dynamic superposition by resonant quantum manipulation\cite{rqm} and investigate its time evolution. Thus we can compute the transmobility of the particle initially prepared in the superposition state. In this case we find that the transmobility reduces to the weighted sum and is non-topological therefore. This can be traced to the fact that the dynamic evolution of the state is generally not periodic. The second way to create a superposition is to add a nonzero mixing matrix elements mixing the bands. We have considered the topological properties of so created static superposition of two states. In this case we investigate in detail how the integer values of the Chern numbers of the superposition states are restored in accordance with the general theory of characteristic classes. We show a general and important property of the matrix element between two topologically distinct phases that it must vanish at some point in parameter space. This allows the topological transitions to happen.

Within the approach of investigation of static superpositions we also analyze the properties of many-band Hamiltonians. If the number of parameters is not large then it is not possible generically to tune the system to the more than two bands crossing point in parameter space. Therefore we conclude that in this case the topological properties of separate bands in terms of the first Chern numbers are sufficient to describe the system completely. More precisely, all the relevant information about the topological properties can be presented in the phase diagram. Those show the quadruple points which is in contrast to usual phase transitions where triple points are common\cite{vol5}. These quadruple points come in two types: the ones that connect 3 or 4 different phases. The Type-II points can be continuously annihilated pairwise or sent to infinity by tuning an additional parameter. The points of the Type-I can in addition to those mechanisms be annihilated by tuning two opposite topological charges in 3-dimensional parameter subspace to the same point. This requires one additional parameter to tune. Therefore, all the crossings are stable with respect to small deviations of the parameters.

Finally, we investigate in detail the phase diagrams for the bilayer Haldane model at some specific choices of parameters. We see the realization of the general features discussed above. In addition, the disappearance of the whole compact region of the topological phase can be achieved by tuning the parameters.


\section{Appendix A: Extended singularities in the bilayer Haldane model \label{Sec:App1}}
In this Appendix we investigate a particular choice of the parameters in Eq.$\eqref{4by4}$ when ${\rm arg}(T_a)= {\rm arg}(T_b)$. This case is somewhat degenerate. In the generic case described in the main text one only has Weyl point singularities in the 3-dimensional space of $q_{1,2}$ and one additional parameter. Those are situated in the corners of the Brillouin zone $K,K^\prime$. Contrary to this, in the degenerate case we find that there is a possibility to have extended 1-dimensional singularities in the 3-dimensional parameter space. Moreover, these can be situated away from the points $K,K^\prime$. We note that the case $|T_a|=|T_b|$ is even more degenerate and opens a possibility to have 2-dimensional singularities in the 3-dimensional parameter space. We do not consider this case here. We also note that the extended singularities are present in the non-generic case, so they can be removed by complicating the model. We report this cone-formation mechanism in this Appendix anyway.

\begin{figure}
	\centerline{\includegraphics[width=0.45\textwidth]{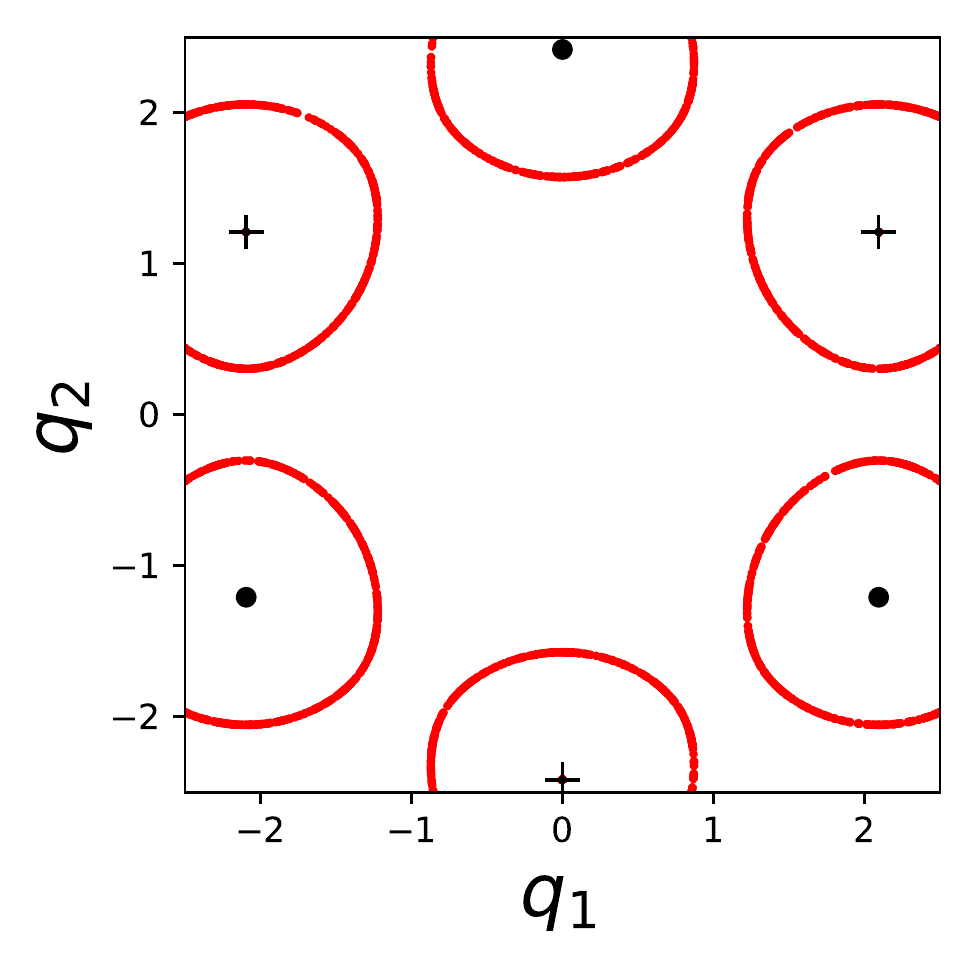}
	}
	\caption{Singular manifolds in the $\vec{q}$-space of the model Eq.$\eqref{4by4}$ in the degenerate case. These are obtained by projection of the Weyl points and extended line singularities from the 3-dimensional space of parameters $q_{1,2}, W$ (see main text) onto the ($(q_1,q_2)$)-plane. The extended singularities can be present in the case when $T_{a,b} \in \Re$. Choice of parameters $-3b^{(1)}=b^{(2)}=-0.3$, $T_a=1$, $T_b=0.5$, $2M^{(1)}=M^{(2)}=2.2$, $\mu=-2$, $W\in [-4;4]$. The projections of the Weyl points are situated in the corners of the Brillouin zone and shown as black dots (point $K$) and black crosses (point $K^\prime$). Extended singularities are shown in red.}
	\label{lines}
\end{figure}

In order to investigate the possibility of the extended singularities away from high-symmetry points $K,K^\prime$ in $\vec{q}$-space for a Hamiltonian $\eqref{4by4}$ it is convenient to rewrite the diagonal blocks $\eqref{2by2}$ as spin Hamiltonians which is always possible for a $2\times 2$ matrix, so
\begin{align}
H_H^{(\eta)}=\vec{B}^{(\eta)}\vec{\sigma}
\label{2by2spin}
\end{align}
where the "magnetic fields" have components $\vec{B}^{(1)}=|B^{(1)}|(\sin \theta \cos \phi,\sin \theta \sin \phi,\cos \theta)$ and $\vec{B}^{(2)}=|B^{(2)}|(\sin \theta^\prime \cos \phi^\prime,\sin \theta^\prime \sin \phi^\prime,\cos \theta^\prime)$. The lengths of $B^{(1,2)}$ and the angles are the functions of initial parameters in $\eqref{2by2}$. The peculiarity of our model is that $\phi=\phi^\prime$ always, which directly follows from $\eqref{2by2}$. One can diagonalize these diagonal blocks in $\eqref{4by4}$ applying a block-diagonal unitary transformation to $\eqref{4by4}$. Upon doing that the upper off-diagonal block is transformed
\begin{widetext}
	\begin{equation}
	{\tilde{\rm T}}=\begin{pmatrix}
	\cos \frac{\theta}{2}\cos \frac{\theta^\prime}{2}T_a+\sin \frac{\theta}{2}\sin \frac{\theta^\prime}{2}T_b & \cos \frac{\theta}{2}\sin \frac{\theta^\prime}{2}T_a-\sin \frac{\theta}{2}\cos \frac{\theta^\prime}{2}T_b\\
	\sin \frac{\theta}{2}\cos \frac{\theta^\prime}{2}T_a-\cos \frac{\theta}{2}\sin \frac{\theta^\prime}{2}T_b & \sin \frac{\theta}{2}\sin \frac{\theta^\prime}{2}T_a+\cos \frac{\theta}{2}\cos \frac{\theta^\prime}{2}T_b
	\end{pmatrix}
	\label{Tangl}
	\end{equation}
\end{widetext}
where we have used that $\phi=\phi^\prime$. We see from $\eqref{Tangl}$ that if the phases of $T_{a,b}$ are different, the matrix elements never vanish, so we always expect avoided crossing away from $K,K^\prime$. If the phases are the same it opens a possibility to set some matrix elements to zero, so the singularities away from high-symmetry points become possible. In fact, if the phases coincide one can gauge them out by another unitary transformation of $\eqref{4by4}$ and bring it to a symmetric form. Due to this additional symmetry there is a possibility to have extended singularities in 3-dimensional parameter space. We indeed see this cone-generation (see Fig.\ref{lines}).

\begin{acknowledgements}
This work was supported by the Netherlands Organization for Scientific Research (NWO/OCW), as part of the Frontiers of Nanoscience program.
\end{acknowledgements}

\bibliographystyle{apsrev4-1}
\bibliography{manuscript}
\end{document}